\documentclass[aps,prl,twocolumn,superscriptaddress,showpacs,amsmath,amssymb]{revtex4}

\usepackage{graphicx}
\usepackage{bm}
\usepackage{amssymb}
\usepackage{hyperref}

\begin{document}

\title{Relativistic Hall Effect}

\author{Konstantin Y. Bliokh}
\affiliation{Advanced Science Institute, RIKEN, Wako-shi, Saitama 351-0198, Japan}
\affiliation{A. Usikov Institute of Radiophysics and Electronics, NASU, Kharkov 61085, Ukraine}

\author{Franco Nori}
\affiliation{Advanced Science Institute, RIKEN, Wako-shi, Saitama 351-0198, Japan}
\affiliation{Physics Department, University of Michigan, Ann Arbor, Michigan 48109-1040, USA}

\begin{abstract}
We consider the relativistic deformation of quantum waves and mechanical bodies carrying intrinsic angular momentum (AM). When observed in a moving reference frame, the centroid of the object undergoes an AM-dependent transverse shift. This is the relativistic analogue of the spin-Hall effect, which occurs in \textit{free space} without any external fields. Remarkably, the shifts of the \textit{geometric} and \textit{energy} centroids differ by a factor of 2, and both centroids are crucial for the Lorentz transformations of the AM tensor. We examine manifestations of the relativistic Hall effect in quantum vortices, and mechanical flywheels, and also discuss various fundamental aspects of this phenomenon. The perfect agreement of quantum and relativistic approaches allows applications at strikingly different scales: from elementary spinning particles, through classical light, to rotating black-holes.
\end{abstract}

\pacs{42.50.Tx, 03.65.Pm, 03.30.+p}

\maketitle

\textit{Introduction.---}
Hall effects represent a group of intriguing phenomena which appear from the interplay between rotation and linear motion of particles. These phenomena are associated with a \textit{transverse drift} of the particle in the direction orthogonal to both its \textit{angular momentum} (AM) and \textit{external force}. For instance, in classical and quantum Hall effects, electrons rotate in a magnetic field and drift orthogonally to the applied electric field \cite{QHE}. In the past decade, various spin Hall effects attracted enormous attention in condensed-matter \cite{SHE}, optical \cite{OHE}, and high-energy \cite{HE} systems. These effects arise from a spin-orbit-type interaction between the intrinsic AM of the particle and its external motion. Despite striking differences between the systems, the spin Hall effects are intimately related to universal AM conservation laws \cite{SHE,OHE,HE,Fedo}. Noteworthily, the intrinsic AM of classical waves or quantum particles can be associated with a circulating internal current which can originate not only from spin but also from \textit{quantum (optical) vortices} \cite{OAM,EV}.

In this Letter, we describe a novel type of Hall effect which naturally arises in special relativity \textit{without} any external fields. We show that the Lorentz space-time transformation of either a rotating mechanical body or a quantum vortex inevitably causes AM-dependent transverse deformations of the object and the Hall shift of its centroid. Moreover, the energy and particle deformations differ by a factor of 2, which is necessary for the correct Lorentz transformations of the AM tensor. Being similar to the spin Hall effect, the phenomenon under discussion is a purely relativistic effect intimately related to the transformations of \textit{time}. There are novel and known examples manifesting the relativistic Hall effect: a moving quantum vortex, a relativistic flywheel \cite{Muller}, and the ``rolling-shutter effect'' that appears on a camera snapshot of a rotating propeller \cite{RSE}. Relativistic transformations of the intrinsic AM can be highly important for classical and quantum waves in moving media \cite{moving}, high-energy physics \cite{Ivanov}, and astrophysics \cite{astro}, where the relative velocities are comparable to the speed of light.

\textit{Relativistic-mechanics approach.---}
It is known that the \textit{shape} of a rigid body and its \textit{intrinsic AM} are invariant upon Galilean transformations in nonrelativistic mechanics. However, they inevitably vary upon Lorentz transformations in special relativity. To begin with, the relativistic AM of a point particle is described by a four-tensor $L^{\alpha \beta }  = r^\alpha   \wedge p^\beta$, where $r^\alpha   = \left( {ct,{\bf r}} \right)$ and $p^\alpha   = \left( {\varepsilon /c,{\bf p}} \right)$ are four-vectors of the coordinates and momentum in the Minkowski space-time \cite{SR}. The antisymmetric AM tensor can be represented by a pair of three-vectors, $L^{\alpha \beta }  = \left( {{\bf H},{\bf L}} \right)$, where ${\bf L} = {\bf r} \times {\bf p}$ is the axial vector of the AM, whereas ${\bf H} = {\bf p}{\kern 1pt} ct - \left( {\varepsilon /c} \right){\bf r}$ is the polar vector marking the rectilinear trajectory of the particle [13]. For a finite-size \textit{body} (i.e., a system of multiple particles), one has to sum the above quantities over all particles. In doing so, ${\bf L} = \sum {{\bf r}_i  \times {\bf p}_i }$ and ${\bf H} = {\bf P}ct - \left( {E/c} \right)\,{\bf R}_E$, where ${\bf P} = \sum {{\bf p}_i }$ and $E = \sum {\varepsilon _i }$ are the total momentum and energy, whereas
\begin{equation}\label{eqn:1}
{\bf R}_E  = \frac{{\sum {\varepsilon _i {\bf r}_i } }}{{\sum {\varepsilon _i } }}~, 
\end{equation}
is the \textit{energy centroid} of the body. The tensor $L^{\alpha \beta }$ is conserved in free space, which includes conservation of the AM ${\bf L}$ and rectilinear motion of the energy centroid according to ${\bf \dot R}_E  = {\bf P}c^2 /E$ \cite{SR}. In the energy-centroid rest frame, ${\bf P} = {\bf R}_E  = {\bf H} = {\bf 0}$ and $E=E_0$.

Let us consider a transformation of the body from the rest frame to a reference frame moving with relativistic velocity ${\bf v}$. For simplicity, we assume that in the rest frame all particles move with nonrelativistic speeds, and in the moving frame they all acquire nearly the same speeds and energy-momentum boosts. This is a relativistic ``paraxial approximation'' for world-lines of the constituent particles, which allows to write one-particle Lorentz transformations for \textit{integral} dynamical characteristics: $E' \simeq \gamma\,E_0$, ${\bf P'} \simeq  - \left( {\gamma E_0 /c^2 } \right){\bf v}$, etc. Here $\gamma  = 1/\sqrt {1 - \left( {v/c} \right)^2 }$ is the Lorentz factor and throughout the paper all quantities in the moving frame are marked by primes. In this manner, applying the Lorentz transformation to the AM tensor with ${\bf H} = {\bf 0}$, we obtain
\begin{eqnarray}\label{eqn:2}
{\bf L'} & = & \gamma \left( {{\bf L}_\bot - \frac{{\bf v}}{c} \times {\bf H}} \right) + {\bf L}_\parallel = \gamma\,{\bf L}_\bot + {\bf L}_\parallel~,\nonumber\\
{\bf H'} & = & \gamma \left( {{\bf H}_\bot + \frac{{\bf v}}{c} \times {\bf L}} \right) + {\bf H}_\parallel = \gamma \frac{{\bf v}}{c} \times {\bf L}~,
\end{eqnarray}
where subscripts $\bot$ and $\parallel$ indicate the vector components orthogonal and parallel to the velocity ${\bf v}$. Since ${\bf L}_\parallel$ is not affected by the transformation, we assume that ${\bf v}\,\bot\,{\bf L}$ and set ${\bf L} = L\,{\bf e}_z$, ${\bf v} = v\,{\bf e}_x$. If one observes the body at $t'=0$, then ${\bf H'} \simeq  - \left( {\gamma E_0 /c} \right)\,{\bf R}_E^{\prime}$, and the transformation (2) yields
%
\begin{equation}\label{eqn:3}
L' = \gamma\,L~,~~~~Y^{\prime}_E  \simeq \frac{v}{{E_0 }}\,L
\end{equation}
Thus, in the moving frame, the AM is enhanced by the relativistic factor of $\gamma$, while the energy centroid experiences a \textit{transverse shift} $Y^{\prime}_E$ proportional to the original AM $L$. This is a manifestation of the \textit{relativistic Hall effect}.

To illustrate this, we consider the example of an axially-symmetric rigid body rotating about the $z$-axis and carrying intrinsic AM ${\bf L} = L\,{\bf e}_z$ in the centroid rest frame (see Fig.~1a). In the moving frame, the body undergoes the Lorentz contraction of the $x$-dimension with the factor of $\gamma^{-1}$, i.e., becomes \textit{elliptical} (Fig.~1b). However, elliptical deformation results in the following change of the intrinsic AM (cf. optical-vortex example \cite{Fedo}):
%
\begin{equation}\label{eqn:4}
L^{\prime\,({\rm int})} = \frac{{\gamma^{-1} + \gamma}}{2}\,L~.
\end{equation}
This follows from the axial symmetry of the body and the equation $L_z  = \sum {x_i\, p_{y_i}  - y_i\, p_{x_i} }$, where $x_i$ experiences contraction with the factor $\gamma^{-1}$, while $p_{x_i}$ grows by the factor $\gamma$. Obviously, the transformation (4) differs from the Lorentz transformation (3). The deficit of AM can be found only in the \textit{extrinsic AM}, produced by the orbital motion of the body centroid, ${\bf R}^{\prime}_C$:
\begin{equation}\label{eqn:5}
L^{\prime\,({\rm ext})} = \left( {{\bf R}^{\prime}_C \times {\bf P'}} \right)_z  = - Y^{\prime}_C \,P'_x~. 
\end{equation}
This situation is quite similar to the \textit{spin Hall effect} in various systems, where variations in the intrinsic AM (spin) are compensated at the expense of the centroid shift generating extrinsic AM \cite{SHE,OHE,Fedo}. Using $P^{\prime}_x \simeq - \left({\gamma E_0/c^2}\right)v$, and requiring $L^{\prime\,({\rm int})}+L^{\prime\,({\rm ext})} = L'$, we arrive at
\begin{equation}\label{eqn:6}
Y^{\prime}_C\;\simeq\;\frac{v}{{2E_0}}\,L~.
\end{equation}
But this shift of the centroid is \textit{two times smaller} than $Y^{\prime}_E$ in Eq. (3), and we again have a contradiction with the Lorentz transformation!
%
\begin{figure}[t]
\includegraphics[width=8.5cm, keepaspectratio]{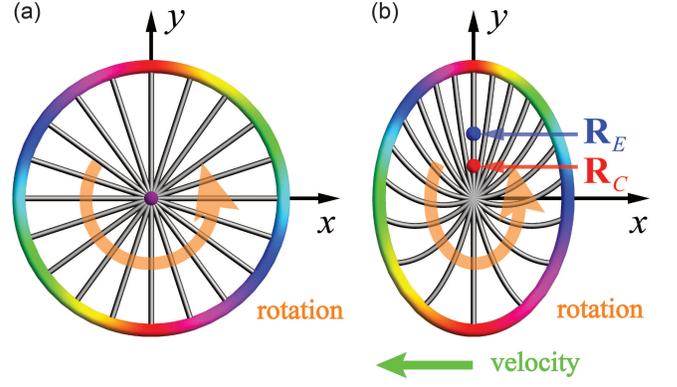}
\caption{(color online). (a) A relativistic flywheel of radius $R$ rotating with angular velocity $\Omega$, $\Omega R/c = 0.7$, in the rest frame \cite{remark I}. (b) Deformations of the wheel shape in the frame moving with velocity $v_x = 0.7\,c$ \cite{Muller}. The dots indicate the positions of the geometric and energy centroids, ${\bf R}^{\prime}_C$ and ${\bf R}^{\prime}_E$.} \label{fig1}
\end{figure}
%

To resolve this discrepancy, note that the extrinsic AM (5) is defined using the \textit{geometric centroid} of the body, determined with respect to the local \textit{number of particles}, $n$, rather than the energy $\varepsilon$:
\begin{equation}\label{eqn:7}
{\bf R}_C  = \frac{{\sum {n_i{\bf r}_i } }}{{\sum {n_i } }}~,
\end{equation}
Assuming that ${\bf R}_E  = {\bf R}_C  = {\bf 0}$ in the rest frame, all transformations become consistent only if the energy and particles centroids differ as
\begin{equation}\label{eqn:8}
{\bf R}^{\prime}_E  =  - {\bf v}\,t' - \frac{{\bf v}\times{\bf L}}{E_0}~,~~{\bf R}^{\prime}_C  =  - {\bf v}\,t' - \frac{{\bf v}\times{\bf L}}{2E_0}~.
\end{equation}

Equation (8) indeed holds true, as it follows from explicit calculations \cite{Muller}, considering the example of a relativistic spinning flywheel. Figure 1 shows numerically-calculated deformations of the wheel in the moving reference frame. Alongside with the $x$-contraction, the $y$-distribution of matter also becomes non-uniform: one can clearly see the crowding and sparseness of the spokes on opposite $y$-sides of the wheel in Fig.~1b. This is as a sort of ``blue'' and ``red'' wavelength shifts in the relativistic Doppler effect because the $y>0$ and $y<0$ sides of the wheel move in opposite directions with respect to the velocity ${\bf v}$. Calculating the density of particles (spokes) along the rim, one can obtain that its centroid is located at $Y_C^{\prime}$, Eq.~(6). However, in addition to the \textit{shape deformations}, a rotating body also acquires \textit{mass deformations}. The $y>0$ and $y<0$ sides of the wheel have different velocities in the moving frame and their constituent particles acquire different local $\gamma$-factors. Owing to this, the $y>0$ particles become \textit{heavier} than the $y<0$ particles. Accounting this yields precisely the two-times-higher transverse shift of the energy centroid, $Y_E^{\prime}$, Eq.~(3) \cite{Muller}. Thus, the \textit{microscopic} picture, Fig.~1, dealing with local particle and energy distributions, complements the \textit{macroscopic} picture, Eqs.~(2)--(6), dealing with the body as a whole, and explains the origin of the different centroid shifts (8).

Importantly, all shape deformations in special relativity originate from the nature of \textit{simultaneity}. Indeed, different values of $x'$ at the instant of time $t'=0$ in the moving frame correspond to different moments $t\neq 0$ in the rest frame. This can be illustrated by representing the Lorentz boost as a rotation in Minkowski space-time. Figure 2a shows a rotating flywheel in the $(x,y)$ plane which propagates freely along the time coordinate $\zeta=ct$ forming a \textit{cylindrical beam} in space-time. In doing so, different points of the wheel have \textit{helical world-lines} within the cylinder. The Lorentz transformation to the moving frame represents a hyperbolic rotation of the coordinates, $Rh(\theta)$: $\left( {x,\zeta } \right) \to \left( {x',\zeta '} \right)$, by the angle $\theta = \tanh^{-1}\!\left( {v/c} \right)$. As a result, the image of the circle at $\zeta'=0$ represents a \textit{tilted cross-section} of the beam, Fig.~2a. This is equivalent to the $x$-\textit{dependent time delay}: condition $\zeta'=0$ yields $\zeta=x\tanh\theta$. It is the combination of this time delay with the circular motion that yields the relativistic Hall effect. Indeed, equidistant spiral world-lines in Fig.~2a become crowded at $y>0$ and sparse at $y<0$ in the tilted cross-section of the cylinder.

Figure~2a also reveals a close similarity between the relativistic Hall effect and recently predicted \textit{geometric spin Hall effect of light} \cite{GSHEL}. The latter also occurs in free space upon observation of a paraxial optical beam with intrinsic AM in the \textit{tilted} reference frame. Let the beam propagate along the $z$-axis, and the tilted frame is obtained via a rotation by an angle $\theta$ about the $y$-axis, $R(\theta)$: $\left( {x,z} \right) \to \left( {x',z'} \right)$. Then, it turns out that the distribution of the \textit{energy flow} through the $z'=0$ plane differs from that through the $z=0$ plane, and its centroid undergoes an AM-dependent transverse shift along the $y$-axis \cite{GSHEL}. This situation is described by the same Fig.~2a, if we assume $\zeta=z$. In this manner, the helical lines represent \textit{streamlines of the energy current} (i.e., the Poynting vector which coils in wave-beams carrying AM \cite{OAM,EV,currents}), and the density of the energy flow through the oblique $z'=0$ plane becomes higher at $y>0$ and lower at $y<0$. Since propagation along the $z$-axis plays the role of time evolution in optics, here one deals with the same ``$x$-dependent time-delay'' effect: $z'=0$ implies $z=x\,\tan\theta$. Moreover, the value of the spin-Hall shift of the energy-flow centroid in the tilted optical beam can be written as \cite{GSHEL} $Y'_{{\rm SHEL}} = (c\tan\theta/2E_0)L$, where $E_0 = \hbar\,\omega_0$ and $L = \hbar\,\ell$ are the photon energy and AM (the integer $\ell$ being the polarization helicity or the vortex charge). For the relativistic Hall effect, $Y_C^{\prime}$, Eq.~(6), has almost identical form with the only change $\tan\theta \to \tanh\theta = v/c$.

The key role of the ``$x$-dependent time-delay'' in the above Hall effects can be exemplified by the so-called ``rolling-shutter effect''. This is a characteristic distortion of the image of a rotating object made by a camera with the shutter moving in the $x$-direction, Fig.~2b \cite{RSE}. Despite the nonrelativistic velocities, the rolling-shutter effect is entirely analogous to the relativistic Hall effect, Fig.~1b.
%
\begin{figure}[t]
\includegraphics[width=8.5cm, keepaspectratio]{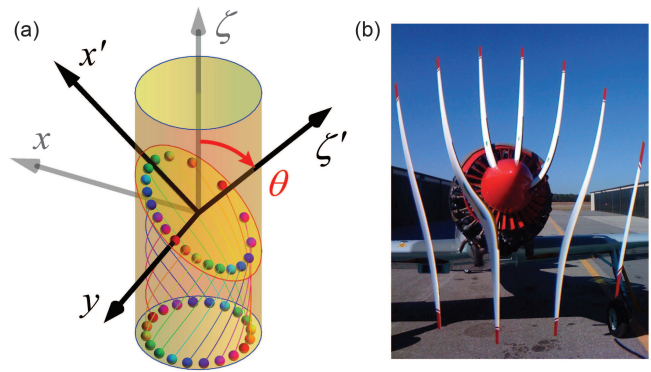}
\caption{(color online). (a) Geometrical explanation of the relativistic Hall effect induced by the Lorentz rotation of the space-time ($\zeta=ct$). The density of helical world-lines of a rotating flywheel becomes asymmetric along the transverse $y$-axis in the moving frame. This also illustrates the geometric spin Hall effect of light produced by the rotation of coordinates ($\zeta=z$) \cite{GSHEL}, where helical streamlines show the energy current in a beam carrying AM. (b) The ``rolling-shutter effect'': a visual distortion and $y$-shift of the centroid of a rotating propeller caused by the $x$-dependent time delay from the moving shutter of the camera \cite{RSE}, cf. Fig.~1b.} \label{fig1}
\end{figure}
%

\textit{Quantum-wave approach.---}
Let us now examine a relativistic quantum (wave) system carrying intrinsic AM. The simplest localized wave carrying intrinsic AM is a \textit{monochromatic Bessel beam} \cite{Ivanov,Bessel} propagating in the $z$-direction and described by the wave function
\begin{equation}\label{eqn:9}
\psi\!\left({\bf r},t\right) \propto J_{\left| \ell \right|}\!\left(\kappa r\right) \exp\left[i\!\left(\ell\,\varphi + k_z z - \omega_0 t\right)\right]~.
\end{equation}
Here $J_n(\xi)$ is a Bessel function, $(r,\varphi,z)$ are cylindrical coordinates, $\kappa$ and $k_z$ are the transverse and longitudinal wave numbers, $\ell = 0, \pm 1, \pm 2,...$ is the AM quantum number, and $\omega_0$ is the frequency. The Bessel beam (9) is an exact solution of the Klein-Gordon relativistic wave equation, if the wave numbers and frequency satisfy the dispersion relation $\left( {\omega _0 /c} \right)^2 - \left( {k_z^2  + \kappa ^2 } \right) = m^2 c^2 /\hbar ^2  \equiv \mu ^2$, with $m$ being the mass of the quantum particles under consideration.

The beam (9) has a cylindrically-symmetric intensity distribution and contains an \textit{optical (quantum) vortex}, i.e., an azimuthal phase $\exp\!\left(i\ell\varphi\right)$ forming a screw phase dislocation along its axis, Fig.~3a \cite{OAM,EV,Nye}. The vortex generates a spiralling energy current in the beam and is intimately related to the well-defined intrinsic AM $L=\hbar\,\ell$ per particle \cite{OAM,EV}. Indeed, using operators of energy, $\hat \varepsilon = i\hbar\,\partial_t$, momentum, ${\bf \hat p} = - i\hbar \nabla$, and AM, ${\bf \hat L} = {\bf \hat r} \times {\bf \hat p}$, one can see that the Bessel beams (9) are eigenfunctions of $\hat \varepsilon$, $\hat{p}_z$, and $\hat L_z  =  - i\hbar\,\partial _\varphi$, with eigenvalues $E_0 = \hbar\,\omega_0$, $P_z  = \hbar\,k_z$, and $L=\hbar\,\ell$.

Since the Klein-Gordon wave equation is Lorentz-invariant, one can find the form of the Bessel beam in the moving reference frame by transforming coordinates $\left( {{\bf r},t} \right)$ in Eq.~(9). 
Substituting the Lorentz transformation 
\begin{equation}\label{eqn:10}
\left( {\begin{array}{*{20}c}
   {ct}  \\
   x  \\
\end{array}} \right) = \gamma \left( {\begin{array}{*{20}c}
   1 & {v/c}  \\
   {v/c} & 1  \\
\end{array}} \right)\left( {\begin{array}{*{20}c}
   {ct'}  \\
   {x'}  \\
\end{array}} \right),~y=y',~z=z',
\end{equation}
into Eq.~(9), the scalar wave function in the moving frame becomes $\psi '\left( {{\bf r'},t'} \right) \equiv \psi \left[ {{\bf r}\left( {{\bf r'},t'} \right),t\left( {{\bf r'},t'} \right)} \right]$. The beam $\psi '\left( {{\bf r'},t'} \right)$ becomes \textit{polychromatic} and moves in the $x'$-direction with velocity $-{\bf v}$. Plotting the intensity $I' = \left| {\psi '\left( {{\bf r'},t'} \right)} \right|^2$ at $t'=0$ one would see an elliptical vortex beam obtained by the Lorentz contraction $x \to \gamma x'$ of the original intensity $I = \left| {\psi \left( {{\bf r},t} \right)} \right|^2$. It is known from optics \cite{Fedo}, that such elliptic beam carries intrinsic AM $L^{\prime\,({\rm int})}$ given by Eq.~(4), which contradicts the Lorentz transformations (2) and (3).

To resolve this discrepancy, note that the ``naive'' wave intensity $I = \left| \psi  \right|^2$ and current ${\bf j} = {\mathop{\rm Im}\nolimits} \left( {\psi^* \nabla \psi } \right)$ do \textit{not} represent the \textit{density of particles} ($\int {I\,} dV$ is not Lorentz-invariant) and the \textit{energy current}. The actual density of particles $I_C$ is given by the zero component of the \textit{probability current}, whereas the energy density $I_E$ and energy current ${\bf j}_E$ are provided by the \textit{stress-energy tensor}. In the case of the Klein-Gordon equation, these quantities read ($\zeta\equiv ct$) \cite{QED}:
\begin{eqnarray}\label{eqn:11}
I_C = - {\mathop{\rm Im}\nolimits} \left[ {\psi ^* \partial _\zeta  \psi } \right],~I_E &=& \frac{1}{2}\left[ {\left| {\partial _\zeta  \psi } \right|^2  + \left| {\nabla \psi } \right|^2  + \mu ^2 \left| \psi  \right|^2 } \right], \nonumber\\
{\bf j}_E &=& - {\mathop{\rm Re}\nolimits} \left[{\left( {\partial _\zeta  \psi } \right)^* \left( {\nabla \psi } \right)} \right].
\end{eqnarray}
For plane waves, this yields simple $\omega$-scalings: $I_E  = \left( {\omega /c} \right)^2 I$, ${\bf j}_E  = \left( {\omega /c} \right)\,{\bf j}$, and $I_C  = \left( {\omega /c} \right) I$, which make no difference for monochromatic beams in the rest frame. However, a Lorentz transformation to the moving frame affects these distributions via local variations of the frequency. Figure~3b shows the transverse distributions of the densities $I^{\prime}_C$ and $I^{\prime}_E$ in the moving beam at $t'=0$. The particle and energy distributions differ from each other and show a characteristic \textit{asymmetry} along the $y$-direction. (This asymmetry is two times higher for the energy distribution because of the $\omega^2$-scaling in $I_E$ versus the $\omega$-scaling in $I_C$.) In the paraxial approximation, $\kappa\ll k_z$, the $y$-asymmetry does not affect the integral energy and momentum of the moving beam \cite{remark II}:
\begin{equation}\label{eqn:12}
E' = \hbar\,\frac{{\int {I^{\prime}_E \,} dV'}}{{\int {I^{\prime}_C \,} dV'}}~,~~{\bf P'} = \hbar\,\frac{{\int {{\bf j}^{\prime}_E \,} dV'}}{{\int {I^{\prime}_C \,} dV'}}~,
\end{equation}
which yield $E'\simeq \gamma E_0$ and ${\bf P'}\simeq P_z {\bf e}_z  - \left( {\gamma E_0 /c^2 } \right){\bf v}$, in agreement with the Lorentz transformations. At the same time, the $y$-asymmetry of the energy and particle densities is crucial in calculations of the AM and the corresponding beam centroids in the moving frame. Indeed, these quantities should be defined as
\begin{equation}\label{eqn:13}
{\bf L'} = \hbar\,\frac{{\int {{\bf r'} \times {\bf j}^{\prime}_E \,} dV'}}{{\int {I^{\prime}_C \,} dV'}}~,~~
{\bf R}^{\prime}_{E,C}  = \frac{{\int {{\bf r'}\,I^{\prime}_{E,C} \,} dV'}}{{\int {I^{\prime}_{E,C} \,} dV'}}~,
\end{equation}
and ${\bf r'}$ in the integrands cuts the \textit{antisymmetric} parts of $I^{\prime}_{C}$, $I^{\prime}_{E}$, and ${\bf j}^{\prime}_E$. Evaluating the integrals (13) for the Bessel beam (9) in the moving frame (10) yields \cite{remark II} $L' \simeq \gamma\,L$, $Y^{\prime}_E  \simeq (v/E_0)L$, and $Y^{\prime}_C \simeq (v/2 E_0)L$, in \textit{exact} correspondence with the mechanical results (3), (6), and (8).
%
\begin{figure}[t]
\includegraphics[width=8.5cm, keepaspectratio]{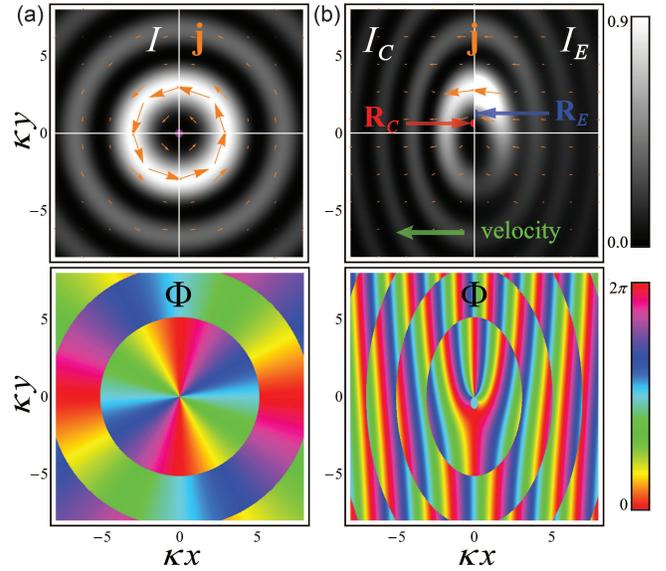}
\caption{(color online). Intensity $I$, current ${\bf j}$, and phase $\Phi=\arg\psi$ distributions for the nonparaxial Bessel beam (9) with $\ell=2$, $k_z c/\omega_0 = 0.33$, $\kappa c/\omega_0 = 0.8$, and $\mu c/\omega_0 = 0.5$ \cite{remark I} in the rest frame (a) and in the frame moving with velocity $v_x=0.8\,c$ (b). In the moving frame, the particle and energy distributions, $I_C$ and $I_E$, acquire centroid shifts ${\bf R}^{\prime}_C$ and ${\bf R}^{\prime}_E$, whereas the charge-2 screw phase dislocation metamorphoses into the moving charge-2 edge dislocation \cite{Nye} (cf. Figs. 1 and 2b).} \label{fig1}
\end{figure}
%

Thus, the deformation of a localized wave carrying intrinsic AM in the moving reference frame is entirely analogous to the deformations of a rigid body in special relativity. Alongside with the distortions of the particle and energy distributions, Fig.~3 demonstrates a remarkable metamorphosis of the phase patterns in the moving frame. A \textit{screw} wavefront dislocation with symmetric radial phase fronts in the rest frame (Fig.~3a) transforms into the moving \textit{edge-screw} wavefront dislocation with crowding of the phase fronts at $y>0$ and sparseness at $y<0$ (Fig.~3b) \cite{Nye}. This is quite similar to the redistribution of the spokes in the relativistic flywheel, cf. Figs.~1 and 2b.

\textit{Discussion.---}
The relativistic Hall effect illuminates fundamental aspects of the AM. It is known that the interplay of relativistic and quantum theories can be a subtle issue which often rises nontrivial questions and paradoxes. Our description makes relativistic and quantum aspects of the AM fully consistent with each other. Furthermore, several fundamental consequences can be immediately deduced from the above theory.

Specifically, the \textit{impossibility to shrink an object carrying intrinsic AM to a point} follows from the relativistic Hall effect [13b]. Since the particle and energy centroids must be within the body boundaries, the minimal radius of the body can be estimated from Eqs.~(3) and (6) at $v\simeq c$ as $R_{\min} \sim c\left| L \right|/E_0$. This estimation works in strikingly contrasting situations. First, substituting $E_0 = \hbar\,\omega_0$ and $L = \hbar\,\ell$, $R_{\min}$ determines the minimal radius of a tightly focused optical beam carrying intrinsic AM \cite{Alonso}. Second, for $E_0=Mc^2$, $R_{\min}$ estimates the minimal Schwarzschild radius of a rotating Kerr black-hole [13a]. Finally, if $E_0=mc^2$ and $L \sim \hbar$, $R_{\min}$ yields the Compton wavelength, i.e., the minimal radius of the Dirac electron wave packet with spin [4c].

In addition, the relativistic Hall effect sheds light on the \textit{spin Hall effect of a Dirac electron} moving in an external potential. The latter effect is caused by the spin-orbit interaction, but the nonrelativistic limit of the covariant Dirac equations of motion predicts the spin-Hall deflection of the electron trajectory \textit{two times larger} than that derived from the standard spin-orbit Hamiltonian \cite{HE}. Since it is the \textit{energy} centroid that follows the one-particle equation of motion, this might yield the factor of 2 in the spin-Hall deflection. Indeed, the covariant equation of motion \cite{HE}, $\delta {\bf \dot r} = {\bf p}/m + ({\bf \dot p}\times{\bf S})/(m^2 c^2)$ (${\bf S}$ being the spin), can be immediately derived differentiating ${\bf R}^{\prime}_E = -{\bf v}t' - ({\bf v}\times{\bf L})/E_0$, with ${\bf L}={\bf S}$, $E_0 = mc^2$, ${\bf p}=-m{\bf v}$, and \textit{without} involving any electromagnetic interactions.

We acknowledge fruitful discussions with Y. P. Bliokh and support from the European Commission (Marie Curie Action), LPS, NSA, ARO, NSF grant No. 0726909, JSPS-RFBR contract No.~09-02-92114, Grant-in-Aid for Scientific Research (S), MEXT Kakenhi on Quantum Cybernetics, and the JSPS through its FIRST program.



\begin{thebibliography}{99}
\bibitem{remark I} To make the relativistic Hall deformations clearly visible in the numerical calculations of Figs.~1 and 3, we take a relativistically-rotating wheel with $R\sim c/\Omega$ and nonparaxial Bessel beam with $\kappa^{-1}\sim c/\omega_0$, which have about the minimal possible radii. These parameters are beyond the ``paraxial approximations'' in use, but all the results are in complete agreement with the paraxial equations.
\bibitem{remark II} Hereafter, the integrals over the whole space diverge for Bessel beams. However, integration over a large enough (but finite) symmetric volume around the beam axis is assumed, which yields a meaningful result independent of this volume. The volume of integration should be Lorentz-contracted upon transition to the moving frame.
\end{thebibliography}
\end{document}